\documentclass[aps,pra,twocolumn,letter]{revtex4-1}
\usepackage[english]{babel}
\usepackage{subeqnarray}
\usepackage{hyperref}
\usepackage{graphicx}
\usepackage{mathrsfs}
\usepackage{bm}
\usepackage{amssymb}
\usepackage{amsmath}
\usepackage{bbold}

\newcommand{\vv}[1]{\mathbf{#1}}
\newcommand{\vb}[1]{\boldsymbol{#1}}
\newcommand{\bra}[1]{\langle#1|}
\newcommand{\ket}[1]{|#1\rangle}
\newcommand{\scalar}[2]{\langle #1 | #2 \rangle}

\newcommand{\eq}[1]{(\ref{eq:#1})}
\newcommand{\fig}[1]{Fig. \ref{fig:#1}}
\newcommand{\abs}[1]{\left|#1\right|}

\begin{document}
\title{Retrieval of multiple spin waves from a weakly excited, metastable atomic ensemble}
\date{\today}
\author{F. Bariani}
\author{T.A.B. Kennedy}
\affiliation{School of Physics, Georgia Institute of Technology, Atlanta, GA, 30332-0430, USA}
\begin{abstract}

The emission of light from a multiply excited atomic ensemble is examined and it is shown how symmetric (spin-wave) and non-symmetric states of excitation radiate into spatially separate field modes. This observation has potential application to single photon generation and spin wave entanglement, since in the presence of atomic interactions it can result in isolated single photon emission into a phase-matched field mode.

\end{abstract}

\maketitle

\section{Introduction}

Single excitation states of matter and light are an important resource for quantum communication and computing protocols. In this paper we consider cold, optically thick atomic ensembles as the medium for the creation of matter excitations and the source of single photons emitted into a well-defined mode \cite{quantum_interface_review}. The DLCZ quantum repeater protocol \cite{DLCZ} employs atomic ensembles as quantum memory elements, interfaced with flying light qubits: the detection of a Raman scattered photon heralds the creation of a matter excitation in the form of a spin wave, and its coherent storage provides a way to realize deterministic single photons through a read-out protocol involving feedback \cite{kennedy09}. To avoid noise due to multiple excitations of the medium, weak laser excitation is employed. This bottleneck limits the speed of single photon production providing motivation for alternative approaches, for example, harnessing atomic interactions in order to limit the deleterious effects of laser induced multiple excitations.

The Rydberg blockade mechanism, in which the excitation of a particular atom prevents excitation of nearby atoms in a small sample, is based on strong dipole-dipole interactions between atomic Rydberg levels \cite{lukin2001,saffman2002,Pedersen2009,saffman_review,pfau2011}. The subsequent decay of the excited atom will generate a single photon, without the need to rely on repeated weak laser excitation cycles. An alternative proposal, operating with a larger sample, outside of the Rydberg blockade regime, involves using Rydberg atom resonant dipole-dipole interactions to decouple the decay of multiply excited atomic configurations from the preferred output radiation mode, leaving only the singly excited component to decay with the emission of a single photon \cite{barianishort}. 
In this paper we discuss this radiative decay mechanism in detail, showing explicitly how interaction induced atom-pair phase shifts decouple multiply excited atomic states from the phase-matched mode. 
As a consequence, we show that the quantum statistics of the phase matched field mode has a strongly single-photon character.
General properties of the emission from collective atomic many body states and their mapping onto the radiation field has been considered in \cite{porras2008}, while the mode structure for thermal ensembles of atoms was discussed in \cite{duan2002}.

The role of radiative interactions between atoms has been studied since Dicke pointed out how the decay of collective excitations may lead to \emph{superradiant} emission \cite{dicke1954,scullyscience2009}.
Other workers investigated temporal and spatial signatures of the emitted radiation pointing out that induced atom-atom coherence results in directional emission of light, according to the shape of the atomic sample \cite{ernst1968,ernst1969,LehmbergPRA1,LehmbergPRA2,eberly1971}.
Both the limiting cases of single excitation \cite{ernst1969} and of a completely inverted medium \cite{ernst1968} have been treated. A review of the theory of collective spontaneous emission and related works can be found in \cite{reviewharoche82}.

The goal of controlling light sources at the single photon level has recently put renewed focus on the treatment of the decay of a single collective excitation stored in an atomic gas \cite{scullyPRL2006,eberly2006,scullylaserphys2007,kurizki2007,bermanPRA2009,bermanPRA2011,piovellaPRA2011,scullyPRL2009,SvidzinskyOC2009,SvidzinskyOC2010,scullypra2010, manassahPRA2010,manassahoptcomm08,manassahoptcomm08bis,manassahoptcomm09}.
The collective Dicke state, a single excitation symmetrically shared among all of the atoms, decays with superradiant character only if it is stored in an atomic cloud with dimensions much smaller than the wavelength of the emitted radiation. The excitation remains trapped in the opposite limit of a large ensemble \cite{scullypra2010}. In this latter case, a different quantum state, often called a symmetric timed (or phased) Dicke state \cite{scullylaserphys2007}, shows fast decay in a given direction. 
In this work, we show that phase matched emission also occurs for multiply excited symmetric timed Dicke states in a large enough atomic ensemble.
When two or more atoms are excited in Rydberg states, however, interactions dephase the atomic synchronization imparted by the external laser fields. As a result their phase-matched emission is suppressed.
The role of virtual processes, neglected in earlier works, has been shown to modify the decay of trapped states \cite{scullypra2010,manassahPRA2010}.
Conditions and protocols required to create such state have been recently investigated \cite{scullyPRL2006,bermanPRA2009}. Furthermore, new systems have been explored which show very interesting light emission properties in novel trapping geometries \cite{lesanovskyPRA2010,lesanovskyPCCP2011}.

The paper is organized as follows. We briefly review the theory of interaction induced spin wave dephasing according to \cite{barianishort} in Section \ref{sec:dephasing}: this discussion motivates our focus on the phase-matching condition for multiple emission processes and the influence of atomic interactions on it.
In Section \ref{sec:hamiltonian}, we discuss the Hamiltonian for the interactions between atom and field in the retrieval process. Section \ref{sec:single} contains the main results of the treatment of the single excitation \cite{scullylaserphys2007,scullyPRL2009} which is briefly reviewed in Appendix \ref{sec:appendixA}. The latter gives context for the analysis of multiple excitations in Section \ref{sec:double} (two excitations) and \ref{sec:n} ($n$ excitations), leading to identification of a phase matched radiation mode onto which the spin-wave properties are mapped. We summarize the results with final remarks in Section \ref{sec:conclusion}. Appendix \ref{sec:appendixB} provides technical details of the Wigner-Weisskopf approximation in the case of multiple atomic excitations.

\section{Interaction induced dephasing of multiple spin waves}
\label{sec:dephasing}
\begin{figure}[htbp]
\begin{center}
\includegraphics[width = \columnwidth]{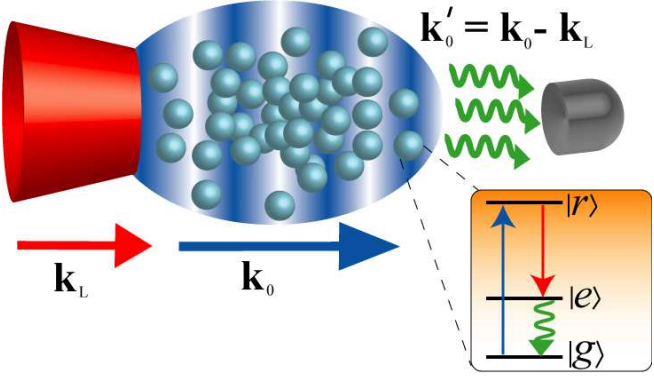}
\caption{Schematic illustration of the retrieval of spin waves (wavevector $\vv{k}_0$) from a large cold atomic ensemble. The excitations are stored in the metastable state $\ket{r}$ (e.g. a Rydberg level): a laser with wavevector  $\vv{k}_L$ is incident on the atomic cloud promoting atoms into the fast decaying state $\ket{e}$. The signal is detected according to a phase-matching condition in the direction given by $\vv{k}_0 - \vv{k}_L$.}
\label{fig:setup}
\end{center}
\end{figure}

We consider a system of $N$ atoms. We focus on three single atom energy levels that are coupled by light fields: a metastable (Rydberg) level $\ket{r}$, an intermediate level $\ket{e}$ and the ground level $\ket{g},$ see \fig{setup}.
The ground state of the atomic ensemble is the product state $\ket{G} = \ket{g_1....g_N}$; from here-on we will write atomic product states listing only those atoms excited out of the single-atom ground state, e.g., $ \ket{s_{\mu_1}...s'_{\mu_n}}$, where $s...s' \in \{e,r\}$ and the indices $(\mu_1,...\mu_n)$ must all be different since the creation of two excitations on the same atom is forbidden. Transitions between different atomic levels are described by the single-particle operators $\hat{\sigma}_{\mu}^{ss'} = \ket{s_{\mu}}\bra{s'_{\mu}}$.
We define collective excitations of level $\ket{r}$ in terms of spin waves, whose destruction operator is given by
\begin{equation}
\hat{S}_{\vv{k}_0} = \frac{1}{\sqrt{N}} \sum_{\mu=1}^{N} e^{i \vv{k}_0\cdot \vv{r}_{\mu}} \hat{\sigma}_{\mu}^{gr},
\end{equation}
with $\vv{r}_{\mu}$ the position of atom $\mu$ and $\vv{k}_0$ the wavevector associated with the excitation. We also refer to this operator as the annihilation operator for symmetric atomic excitations that in Sections \ref{sec:single}, \ref{sec:double} and \ref{sec:n}, we relate with the symmetric states in the timed-Dicke basis.
We consider atomic motion to be frozen, an approximation that requires the wavelength of the stored spin waves to be longer than the distance traveled by an atom during the storage and retrieval time \cite{kennedy09}.

In order to describe the interaction induced dephasing of multiple spin waves, we consider a simple model with up to two excitations. Assume the initial excitation process brings the atomic ensemble to the state:
\begin{equation}
\ket{\Psi_0} = c_0 \ket{G} + c_1 \hat{S}^{\dagger}_{\vv{k}_0} \ket{G} +  c_2 \frac{(\hat{S}^{\dagger}_{\vv{k}_0})^2}{\sqrt{2}} \ket{G}.
\label{eq:Psi0}
\end{equation}
Two-body interactions between atoms excited to level $\ket{r}$, that take place after the excitation for a time $T$, lead to a phase shift
\begin{equation}
\ket{r_{\mu} r_{\nu}} \rightarrow e^{i \Phi_{\mu\nu}} \ket{r_{\mu} r_{\nu}},
\label{eq:phase_shift}
\end{equation}
where $\Phi_{\mu\nu} = \mathcal{U}_{\mu\nu} T/\hbar$  is proportional to the two-body interaction strength $\mathcal{U}_{\mu\nu}$.
A signature of the multiparticle dephasing may be found in the two-particle spin wave correlation function defined by
\begin{equation}
g^{(2)} = \frac{\langle \hat{S}_{\vv{k}_0}^{\dagger} \hat{S}_{\vv{k}_0}^{\dagger} \hat{S}_{\vv{k}_0}\hat{S}_{\vv{k}_0} \rangle}{\langle\hat{S}_{\vv{k}_0}^{\dagger} \hat{S}_{\vv{k}_0}\rangle^2}.
\label{eq:g2}
\end{equation}
For a single excitation $g^{(2)} = 0.$
To illustrate the effect of dephasing, we calculate \eq{g2} for the state resulting from \eq{Psi0} after the phase shifts \eq{phase_shift}
\begin{equation}
g^{(2)}(T) = \frac{\abs{c_2}^2 \abs{ \frac{\sqrt{2}}{N^2} \sum_{\mu,\nu}  e^{i\Phi_{\mu\nu}}}^2}{ \left[ \abs{c_1}^2  +  \abs{c_2}^2  \frac{2}{N^3}  \sum_{\mu} \abs{\sum_{\nu} e^{i\Phi_{\mu\nu}}}^2\right]^2}.
\label{eq:g2dephasing}
\end{equation}
For \eq{Psi0}, a truncated coherent state $c_n = 1/\sqrt{e\,n!}$, $n=~0,1,2$, with $\Phi_{\mu\nu} \rightarrow 0$,
\begin{equation}
g^{(2)}(0) = \frac{e}{4} < 1,
\end{equation}
whereas for a coherent state, $c_n = 1/\sqrt{e\,n!}$, $n=~0...\infty$, $g^{(2)}(0) =1$.
For a random distribution of phase shifts, and a sufficiently long interval $T > \tau$, the sums in the numerator and denominator of \eq{g2dephasing} will vanish due to destructive interference of the complex amplitudes.
Since the denominator contains a constant term, we may take the approximation
\begin{align}
g^{(2)} &\xrightarrow{T > \tau}  \frac{2 \abs{c_2}^2}{\abs{c_1}^4}  \abs{\frac{1}{N^2} \sum_{\mu\nu}  e^{i\Phi_{\mu\nu}}}^2  \nonumber \\
&=  4g^{(2)}(0) \abs{\frac{1}{N^2} \sum_{\mu\nu}  e^{i\Phi_{\mu\nu}}}^2.
\end{align}
The interactions can act to suppress the two-particle correlations as though only single excitations were present. The result is potentially useful as a source of single photons if the spin-wave mode can be mapped onto a well-defined radiation field mode determined by $\vv{k}_0$, and if the dephased multiple excitations do not couple to this same mode in the radiative emission process. The following sections are devoted to analyzing these two issues, by identifying the states involved in the laser excitation, dephasing and retrieval processes and their radiative decay channels.

\section{Description of light-atoms coupling in the retrieval process}
\label{sec:hamiltonian}
The retrieval process is sketched in \fig{setup}. A classical laser pulse couples to the $\ket{r} - \ket{e}$ transition and it is characterized by a Rabi frequency $\Omega_L$, wavevector $\vv{k}_L$ and angular frequency $\omega_L$, while the states $\ket{e}$ and $\ket{g}$ are coupled to the continuum of quantized electromagnetic (e.m.) modes. We label these modes with the index $\phi = (\vv{k},\lambda)$, where $\vv{k}$ and $\lambda$ designate the wavevector and the polarization, respectively; the energy of the photon is $\hbar \omega_k = \hbar c k$, where $c$ is the speed of light and $k = \abs{\vv{k}}$. The general state of the field, containing $m$ photons $(\phi_1....\phi_m)$ in $p$ different modes, is given by $\ket{\Psi_{e.m.}} =  \ket{\phi_1....\phi_m}= \hat{a}^{\dagger}_{\phi_1}.....\hat{a}^{\dagger}_{\phi_m} \ket{0}/\sqrt{\epsilon^{\phi_1....\phi_m}}$, where the operator $\hat{a}^{\dagger}_{\phi}$ is the creation operator for photons in mode $\phi$ and the factor $\epsilon^{\phi_1....\phi_m} = \prod_{i=1}^{p} (n_{i}!)$ introduces the correct normalization depending on the populations of the different modes \cite{ernst1968}. The state $\ket{0}$ is the vacuum state defined by $\hat{a}_{\phi} \ket{0} = 0, \forall \phi$.

The Hamiltonian for the coupled radiation field and atoms, in the eletric dipole approximation, is given by $\hat{H} = \hat{H}_0 + \hat{V}$, where:
\begin{align}
& \hat{H}_0  = \sum_{\mu = 1}^{N} \sum_{s = g,e,r}^{} \hbar \omega_{s} \hat{\sigma}_{\mu}^{ss} + \sum_{\phi} \hbar \omega_{k} \hat{a}^{\dagger}_{\phi} \hat{a}_{\phi}, \\
& \hat{V} = \frac{\hbar \Omega_L}{2} \sum_{\mu=1}^{N} \left[ e^{i(\vv{k}_L \cdot \vv{r}_{\mu} - \omega_L t)} \hat{\sigma}^{re}_{\mu} + h.c. \right]  \nonumber \\
    &     -  \sum_{\mu=1}^{N} \sum_{\phi} \left[i (\mathcal{E}_{k} \vb{\epsilon}_{\phi} \hat{a}_{\phi} e^{i \vv{k} \cdot \vv{r}_{\mu}}) \cdot (d_{eg} \vv{n}_{\mu} \hat{\sigma}^{eg}_{\mu}) + h.c. \right] \label{eq:potential} .
\end{align}
We have defined the electric field per photon $\mathcal{E}_{k} = \sqrt{\hbar\omega_k / (2 \epsilon_0 V)}$. The polarization vector for mode $\phi$ is $\vb{\epsilon}_{\phi}$ while $d_{eg}\vv{n}_{\mu}$ is the electric dipole matrix element for the transition $\ket{e}$-$\ket{g}$ of atom $\mu$. In the following, we introduce the radiation-matter coupling constant $\hbar g_{\phi\mu} = - \mathcal{E}_k d_{eg} (\vb{\epsilon}_{\phi} \cdot \vv{n}_{\mu})$.
We make the rotating wave approximation (RWA) for both transitions. For the $\ket{r}-\ket{e}$ coupling, RWA holds because we consider a resonant or quasi-resonant laser; for the quantized field, it has been shown that virtual processes have only a minor effect on fast decaying states, which are the focus of the present work \cite{scullyPRL2009,scullypra2010}.
For this reason we make the RWA allowing to simplify the problem as the total number of excitations in the system, atomic plus photonic, is conserved. In the following sections, we consider separately the dynamics of states with a different number of excitations.

\begin{figure*}[htbp]
\begin{center}
\includegraphics[width = \textwidth]{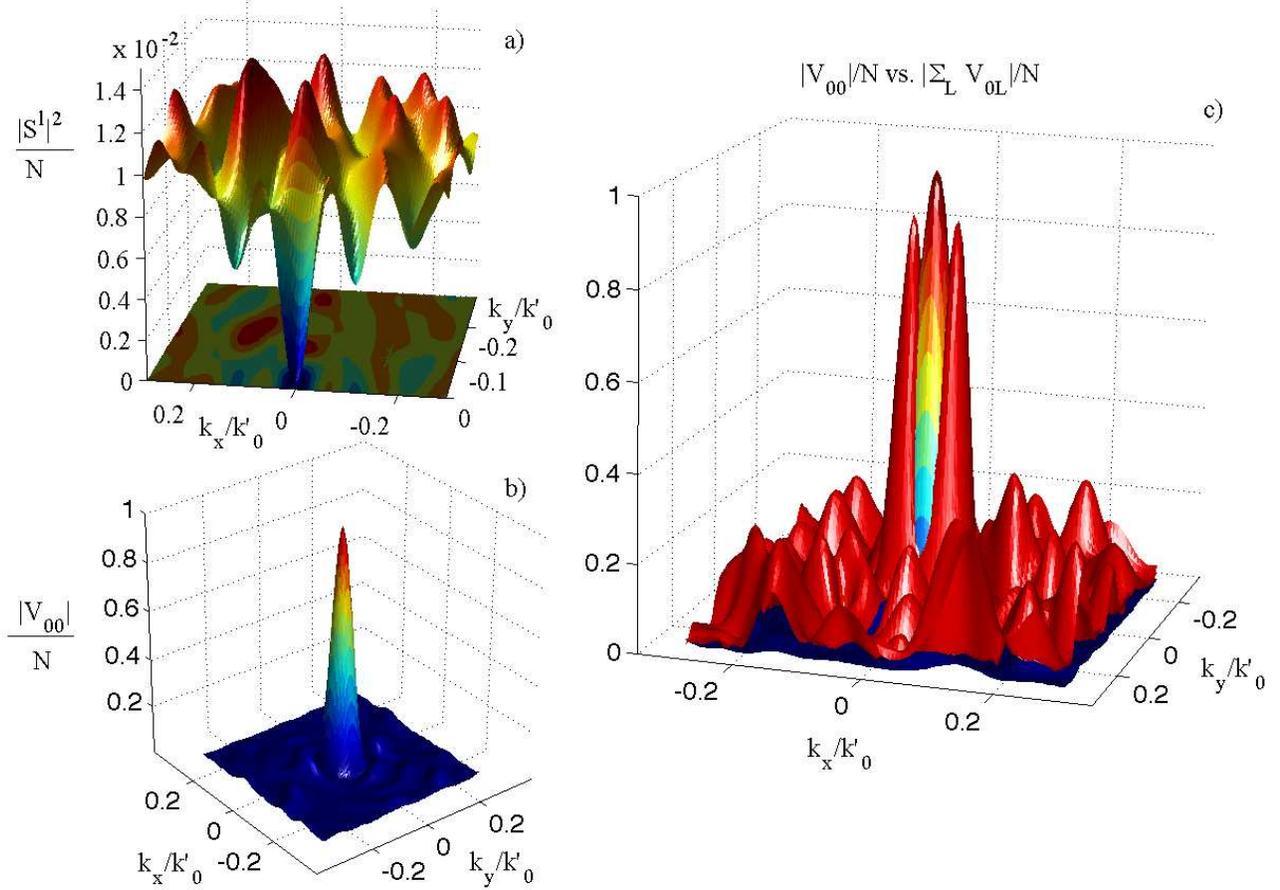}
\caption{(Colors online) Comparison of coupling strengths for the decay of a single spin wave into symmetric and non-symmetric modes from Eq.~\eq{decayE0}. Cubic sample, side $L = 10 \mu$m, $N = 100$ particles $(\rho = 10^{11} \mathrm{cm}^{-3})$. a) Coupling between a spin wave and one of the non-symmetric Dicke state ($\ell = 99$): $S^{\ell} = \sum_{\gamma} S^{\ell}_{\gamma}/\sqrt{\ell(\ell+1)}$ according to \eq{Sfunction1}. b) Normalized coupling of the spin wave to itself, $V^{[1,1]}_{00}(\vv{k})$, see Eq.~\eq{V0011}. c) Comparison of normalized $V^{[1,1]}_{00}(\vv{k})$ (colormap) with the total coupling to non-symmetric modes $\sum_{\ell} V^{[1,1]}_{0\ell}(\vv{k})$ (red surface) : we assume equal amplitudes for all the modes. We assume $\vv{k}'_0 \parallel \hat{z}$. The colormaps are based on the minimum (blue) and maximum (red) values of each plot.}
\label{fig:mode_coupling}
\end{center}
\end{figure*}

\section{Single Excitation}
\label{sec:single}
The theory of a single excitation of the atomic ensemble \cite{scullylaserphys2007,scullyPRL2009} is discussed in appendix \ref{sec:appendixA}. Here, we summarize the most relevant results.
Using the notation previously defined, the state vector may be written using a product state atoms-field basis as,
\begin{equation}
\ket{\Psi}  =    \sum_{\phi} G^{\phi} \ket{G;\phi} + \sum_{\mu} \left( E_{\mu} \ket{e_{\mu};0} + R_{\mu} \ket{r_{\mu};0}\right).
\label{eq:single_state}
\end{equation}
Applying the Wigner-Weisskopf (WW) approximation \cite{scullyquantopt} to the equations of motion \eq{single_exc}, we derive a system of equations for the amplitudes of the excitations in level $\ket{e}$:
 \begin{align}
\frac{\partial E_{\mu}}{\partial t} =& - \frac{\Omega_L(t)}{4} \int_0^{t} dt' e^{i\Delta\omega_L\tau} \Omega_L(t') E_{\mu}(t') \nonumber \\
&- \frac{\Gamma}{2} \sum_{\nu = 1}^N f^{k_{eg}}_{\mu\nu} E_{\nu}(t),
\label{eq:single_decay}
\end{align}
where $\Gamma = \omega^3_{eg} d^2_{eg}/(3 \pi \epsilon_0 \hbar c^3)$ is the single-particle decay rate for the excited level $\ket{e}$. We have defined $\Delta\omega_k = \omega_{eg} - \omega_{k}$, $\Delta\omega_L = \omega_{re} - \omega_{L}$ and $\tau = t' - t$. We define $\omega_{ss'} = \omega_s - \omega_{s'}$.
The function $f$ is defined in \eq{f_function} according to ref.~\cite{LehmbergPRA1}.

 Employing the timed Dicke basis \eq{scully_single} rather than the atomic product states as above, leads to equations for the amplitudes $\mathscr{E}_{\ell}$ defined in \ref{sec:appendixA}. In particular, for the symmetric state $(\ell = 0)$, the unique state in which all atoms share the excitation with equal probability, we obtain
\begin{widetext}
\begin{align}
\frac{\partial \mathscr{E}_{0}}{\partial t} =     \frac{-i \Omega_L}{2} e^{-i\Delta\omega_L t}  \mathscr{R}_{0} - \sum_{\phi} g^2_{\phi} \int_{0}^{t}dt' e^{-i\Delta\omega_k \tau} & \left[ V^{[1,1]}_{00}(\vv{k}) \mathscr{E}_0 + \sum_{\ell=1}^{N-1} V^{[1,1]}_{0\ell}(\vv{k}) \mathscr{E}_{\ell}  \right].
\label{eq:decayE0}
\end{align}
\end{widetext}
We have introduced coupling functions, $V^{[1,1]}_{0 \ell}(\vv{k}),$ $\ell = 0,1,2,...,$ all of which depend implicitly on the phase matched direction $\vv{k}'_0 = \vv{k}_0 - \vv{k}_L$,
\begin{align}
&V^{[1,1]}_{00}(\vv{k}) = \sum_{\mu,\nu} \frac{1}{N} e^{i (\vv{k} - \vv{k}_0')\cdot (\vv{r}_{\mu} - \vv{r}_{\nu})}, \label{eq:V0011}\\
&V^{[1,1]}_{0\ell}(\vv{k}) = \mathcal{C}_{N\mathcal{L}} \sum_{\mu=1}^{N} e^{i(\vv{k}- \vv{k}_0')\cdot \vv{r}_{\mu}}    \sum_{\beta = 1}^{\ell} S^{\ell}_{\beta} (\vv{k} - \vv{k}'_0) \label{eq:V0ell11}
\end{align}
where
\begin{equation}
S^{\ell}_{\beta} (\vv{k} - \vv{k}'_0) =  \left(e^{i (\vv{k}_0' - \vv{k})\cdot \vv{r}_{\beta}} - e^{i (\vv{k}_0' - \vv{k})\cdot \vv{r}_{\ell + 1}} \right).
\label{eq:Sfunction1}
\end{equation}
The normalization coefficient is $\mathcal{C}_{N\mathcal{L}} = 1/\sqrt{N\ell(\ell+1)}.$
Some illustrations of the coupling functions are shown in \fig{mode_coupling}.

The expression \eq{V0011} is peaked for $\vv{k} = \vv{k}'_0$ and decreases far from this condition due to destructive interference of the different atomic phases. In the limit of a large number of atoms, it can be thought of as a non-zero width Dirac delta function.
As pointed out in \cite{scullyPRL2009}, the coupling \eq{V0ell11} is the product of two terms. The first sum is the same as in \eq{V0011} and is peaked in the vicinity of $\vv{k} = \vv{k}'_0$, while all the $S^{\ell}_{\beta}$ functions are strongly suppressed in that region of the wavevector space, as it is evident from \eq{Sfunction1}.
The result is that in the equation of motion \eq{decayE0} the self-coupling of the symmetric excitation dominates over coupling to the other, non-symmetric, basis states. Therefore, we omit these terms and obtain an equation for the decay of the symmetric state with the generalized decay rate \cite{LehmbergPRA1}
\begin{equation}
\Gamma_N = \Gamma \left(1 + \frac{1}{N} \sum_{\mu}\sum_{\nu\neq\mu} e^{-i \vv{k}_0' \cdot(\vv{r}_{\mu} - \vv{r}_{\nu})} f^{k_{eg}}_{\mu\nu} \right).
\label{eq:gamma_N}
\end{equation}
This quantity is in general complex and contains both the superradiant broadening, $\textrm{Re}(\Gamma_N)$, and (Lorentz-Lorenz) frequency shift of the atomic transition, $\textrm{Im}(\Gamma_N)$.

\begin{figure*}[htbp]
\begin{center}
\includegraphics[width = \textwidth]{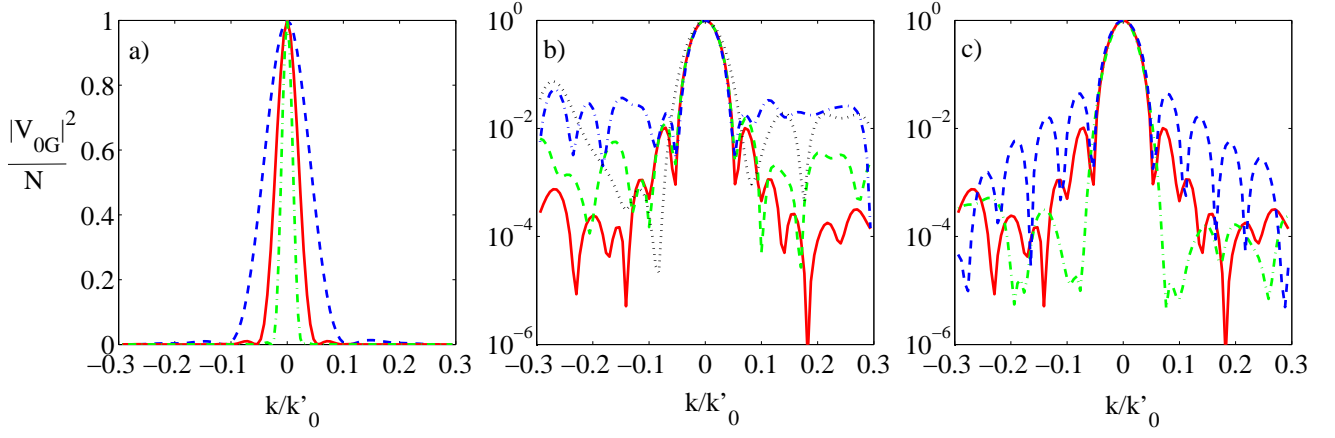}
\caption{(Colors online) Behavior of the normalized emission amplitude $V_{0G}(\vv{k})$ in the phase matched mode. We consider $\vv{k} = (k,0,\sqrt{k_0^{\prime2} - k^2})$ and $\vv{k}'_0  = (0,0,k'_0)$.  a) Effect of diffraction from spherical ensembles with different radius $R$ at fixed density $\rho = 10^{12} \mathrm{cm}^{-3}$. Green dot-dashed line $R = 20 \mu$m, Red solid line $R = 10 \mu$m, Blue dashed line $R = 5 \mu$m.
b) Dependence of the peak on the density of the atoms, for a spherical cloud, $R = 10 \mu$m. Red solid line $\rho = 10^{12} \mathrm{cm}^{-3}$, Green dashed line $\rho = 10^{11} \mathrm{cm}^{-3}$, Blue dot-dashed line $\rho = 10^{10} \mathrm{cm}^{-3}$; for the former two cases the peak is completely symmetric with respect $\hat{x}$ and $\hat{y}$ directions while for the latter density, we also plot the case $\vv{k} = (0,k,\sqrt{k_0^{\prime2} - k^2})$ which is the Black dotted line. This shows the effect of fluctuations at low density. c) Signature of the shape of the atomic cloud in the tails of the phase matched emission. Density is fixed $\rho = 10^{12} \mathrm{cm}^{-3}$. Red solid line is a spherical ensemble, $R = 10 \mu$m, Blue dashed line is a cubic ensemble, side $L = 20 \mu$m, and Green dot-dashed line is a spherically symmetric Gaussian distribution with $\sigma_{x,y,z} = 2.5 \mu$m.}
\label{fig:V00tris}
\end{center}
\end{figure*}

Recall that retrieval of the spin wave stored in the metastable (Rydberg) state $\ket{r}$ is done by means of a single-particle laser $\pi$-pulse, with wave vector $\mathbf{k}_L$ and duration $T$, followed by radiative decay from state $\ket{e}$. In the case of a fast pulse, $T \ll 1/\textrm{Re}(\Gamma_N)$, the probability amplitude for field mode $\phi$ is given by,
 \begin{equation}
G^{\phi}(t)  =  g_{\phi} \frac{e^{-(\Gamma_N/2 + i \Delta\omega_k)t} - 1}{\Gamma_N/2 + i \Delta\omega_k} V_{0G}(\vv{k}).
 \label{eq:mode_ampl}
\end{equation}
We have defined
\begin{flalign}
V_{0G}(\vv{k}) = \sum_{\mu} e^{i \vv{k} \cdot \vv{r}_{\mu}} \bra{\mathsf{E}_0}\hat{\sigma}^{eg}_{\mu}\ket{G} = \frac{1}{\sqrt{N}} \sum_{\mu} e^{i (\vv{k} - \vv{k}_0')\cdot \vv{r}_{\mu}},
\label{eq:V0G}
\end{flalign}
which is a Fourier component of the transition amplitude between the symmetric timed Dicke state $\ket{\mathsf{E}_0}$ and the atomic ground state. The behavior of this term is illustrated in \fig{V00tris}.
Equations \eq{mode_ampl} and\eq{V0G} define a \emph{phase matched mode} \cite{scullylaserphys2007,duan2002,porras2008} into which the emission of light is concentrated.
The non normalized creation operator for this phase matched mode is given by,
\begin{equation}
\hat{b}^{\dagger}_{\vv{k}'_0} = -\sum_{\phi}  g_{\phi} \frac{ V_{0G}(\vv{k})}{\frac{\Gamma_N}{2} + i \Delta\omega_k} \hat{a}^{\dagger}_{\phi},
\label{eq:phase_matched}
\end{equation}
with a Lorentzian distribution of emission frequencies \eq{mode_ampl}, of width $\Gamma_N$.
The information stored in the spin wave is retrieved in the direction given by the phase-matching condition $\vv{k} = \vv{k}'_0$.
We can summarize the excitation and retrieval in a simple sequence:
\begin{equation}
\mathscr{R}_0 \xrightarrow{\Omega_L e^{-i \vv{k}_L\vv{r}}} \mathscr{E}_0 \xrightarrow{\Gamma_N, \delta(\vv{k} - \vv{k'_0})} G^{(\vv{k}'_0,\lambda)}.
\end{equation}

\section{Double excitation}
\label{sec:double}

Before treating the general case of an arbitrary number $n$ of excitations (where $n << N$), we study double excitations in detail. This contains the essential arguments of the general case shorn of the heavier book-keeping.

The equations of motion for the three-level atomic structure with two excitations are much more complicated than the single excitation case.
In principle, we should take into account all the possible combinations of two excitations.
However, as discussed in the single excitation case, and in appendix \ref{sec:appendixA}, it is possible to separate the excitation of the metastable state $|r\rangle$ with the $\pi$-pulse from the successive decay of the excitations from the state $\ket{e}$. Furthermore, the Hamiltonian for the laser coupling is symmetric with respect to different atoms and it preserves the symmetry of the initial state. With these observations, we may reduce the problem to the pair of states $\ket{e}$ and $\ket{g}$.

We first write the doubly-excited state in the atomic product state basis as
\begin{align}
\ket{\Psi} = &\sum_{\phi,\phi'} \frac{ \epsilon^{\phi\phi'}}{2} G^{\phi\phi'} \ket{G;\phi\phi'} +  \sum_{\phi,\mu} E^{\phi}_{\mu} \ket{e_{\mu};\phi}  \nonumber \\
 &+ \sum_{(\nu,\mu)} \frac{E_{\mu\nu}}{2} \ket{e_{\mu}e_{\nu};0}.
 \label{eq:double_state}
\end{align}
Here, the normalization factors are chosen so that the summations run independently over all allowed values for the atomic and electromagnetic labels. We use the notation $(\nu,\mu)$ to indicate a sum on both indices with the exclusion of the term $\nu = \mu$, in order to distinguish from the notation $\nu \neq \mu$ that indicates a sum over the first index, different from a given value of the second one.
The equations of motion for the amplitudes are:
\begin{subequations}
\label{eq:double_eq_mot}
\begin{align}
 \frac{\partial E_{\mu\nu}}{\partial t}  = &\;\sum_{\phi} g_{\phi} e^{i \Delta\omega_k t} \left[e^{i \vv{k}\cdot \vv{r}_{\mu}}  E^{\phi}_{\nu} + e^{i \vv{k} \cdot\vv{r}_{\nu}} E^{\phi}_{\mu} \right] \label{eq:double_atoms} \\
\frac{\partial E^{\phi}_{\mu}}{\partial t}  =& \; - \sum_{\nu\neq \mu} \left[ g_{\phi} e^{- i \vv{k}\cdot \vv{r}_{\nu}} e^{-i \Delta\omega_k t} E_{\mu\nu}\right]  \nonumber \\
& + \sum_{\phi'} \left[ g_{\phi'} e^{i \vv{k}' \cdot\vv{r}_{\mu}} e^{i \Delta\omega_{k'} t} \sqrt{\epsilon^{\phi\phi'}} G^{\phi \phi'} \right] \label{eq:atom_photon}\\
\frac{\partial G^{\phi \phi'} }{\partial t}  =&\; - \frac{1}{\sqrt{\epsilon^{\phi\phi'}}} \sum_{\mu} \left[g_{\phi} e^{-i \vv{k}\cdot \vv{r}_{\mu}} e^{-i \Delta\omega_k t} E^{\phi'}_{\mu} \right. \nonumber \\
& \left. + g_{\phi'} e^{-i \vv{k}'\cdot \vv{r}_{\mu}} e^{-i \Delta\omega_{k'} t} E^{\phi}_{\mu} \right]
\label{eq:double_exc}
\end{align}
 \end{subequations}

Since we assume that the system is initially loaded with two atomic excitations, we start our analysis from \eq{double_atoms} and \eq{atom_photon}. According to the discussion in appendix \ref{sec:appendixB}, we may neglect the coupling terms depending on $G^{\phi\phi'}$.
We separate the process of emission of the photons in two steps: from the double atomic excitation to a single excitation and then to a two-photon state.
The equations of motion for double atomic excitations are
\begin{align}
\frac{\partial E_{\mu\nu}}{\partial t}  = & - \sum_{\phi} g^2_{\phi} \int_{0}^t dt' e^{-i \Delta\omega_k \tau} \left[\sum_{\sigma \neq \nu} e^{i \vv{k}\cdot (\vv{r}_{\mu} - \vv{r}_{\sigma})} E_{\sigma\nu}(t')  \right. \nonumber \\
& \left. + \sum_{\sigma \neq \mu} e^{i \vv{k}\cdot (\vv{r}_{\nu} - \vv{r}_{\sigma})} E_{\sigma\mu}(t') \right].
\label{eq:23}
\end{align}
This set of equations is the analogue for two excitations of \eq{WW} for the single excitation: the amplitudes of all the possible pairs are coupled together through the continuum of field modes.
In this case, we have suppressed the coupling to the laser field, as discussed previously.
A way to proceed at this point would be to apply the WW approach and then diagonalize the resulting system of equations. As we are interested in symmetric states, however, we choose to investigate the couplings between symmetric and non-symmetric collective atomic excitations, following the treatment of the single excitation.

\subsection{Timed Dicke basis for double excitation}
We introduce the timed Dicke basis for the atomic states. For a single excitation we use \eq{scully_single}, while for the double excitations, the basis is given by:
\begin{subequations}
\begin{flalign}
\ket{\mathsf{E}_{0[2]}} =&\; \sqrt{\frac{1}{\mathcal{N}}} \frac{1}{2} \sum_{(\nu,\mu)=1}^N e^{i \vv{k}'_0\cdot (\vv{r}_{\mu} + \vv{r}_{\nu})} \ket{e_{\mu}e_{\nu}}, \\
\ket{\mathsf{E}_{\ell[2]}} =&\; \frac{1}{\sqrt{\mathcal{L}}} \sum_{\gamma = 1}^{\ell} \left[e^{i \vv{k}'_0 \cdot(\vv{r}_{\gamma(1)} + \vv{r}_{\gamma(2)})} \ket{e_{\gamma(1)}e_{\gamma(2)}} \right. \nonumber \\
& \left. - e^{i \vv{k}'_0\cdot (\vv{r}_{\ell+1(1)} + \vv{r}_{\ell+1(2)})} \ket{e_{\ell+1(1)}e_{\ell+1(2)}} \right].
\end{flalign}
\label{eq:scully_double}
\end{subequations}
Here $\ell \in [1,\mathcal{N} - 1]$,  with $\mathcal{N} = \binom{N}{2}$ . The labels $\ell$ and $\gamma$ are used to label pairs of atoms; $\ell(1)$ and $\ell(2)$ indicates the first and second atoms in the pair; similarly for $\gamma(1)$ and $\gamma(2)$ \footnote{There are several ways to decide how to order the different pairs, but this ordering is not crucial in what follows. For example, The pair $\ell = 1$ may contain the atoms $1$ and $2$, and the pair $\ell = 2$, the atoms $1$ and $3$. In this case we have $1(1) = 1$, $1(2) = 2$, $2(1) = 1$ and $2(2) = 3$.}. The subscript $[2]$ in square brackets refers instead to a two atom excitation state.

We rewrite the state \eq{double_state} in the timed Dicke basis as
\begin{align}
\ket{\Psi} = &\sum_{\phi,\phi'} \frac{ \epsilon^{\phi\phi'}}{2} G^{\phi\phi'} \ket{G;\phi\phi'} + \sum_{\phi} \sum_{\ell=0}^{N-1} \mathscr{E}^{\phi}_{\ell} \ket{\mathsf{E}_{\ell};\phi} + \nonumber \\
 &+ \sum_{\ell = 0}^{\mathcal{N}-1} \mathscr{E}_{\ell[2]} \ket{\mathsf{E}_{\ell[2]};0}.
\end{align}
We derive the equations of motion for the new amplitudes. As above, we integrate the equation for $\mathscr{E}^{\phi}_{\ell}(t)$ and we substitute the result into the equation for $\mathscr{E}_{\ell[2]}$ after dropping the contribution from $G^{\phi\phi'}$:
\begin{align}
\frac{\partial \mathscr{E}_{\ell[2]}}{\partial t}
= &- \sum_{\phi} g^2_{\phi}  \sum_{\ell',\jmath} \sum_{\mu,\nu} e^{i \vv{k} \cdot(\vv{r}_{\mu} - \vv{r}_{\nu})} \bra{\mathsf{E}_{\ell[2]}} \hat{\sigma}^{eg}_{\mu} \ket{\mathsf{E}_{\jmath}} \nonumber \\
& \times   \bra{\mathsf{E}_{\jmath}} \hat{\sigma}^{ge}_{\nu} \ket{\mathsf{E}_{\ell'[2]}} \int_0^t dt' e^{-i \Delta\omega_{k} \tau} \mathscr{E}_{\ell'[2]} (t').
\label{eq:symmetric_double}
\end{align}

The structure of the coupling coefficients suggests separated decay channels for symmetric and non-symmetric states. We define
\begin{equation}
V^{[2,2]}_{\ell\ell'}(\vv{k}) = \sum_{\mu,\nu} e^{ i  \vv{k}\cdot (\vv{r}_{\mu} - \vv{r}_{\nu})}  \bra{\mathsf{E}_{\ell[2]}} \hat{\sigma}^{eg}_{\mu} \hat{\sigma}^{ge}_{\nu} \ket{\mathsf{E}_{\ell'[2]} },
\end{equation}
and calculate the couplings involving at least one symmetric state,
\begin{align}
& V^{[2,2]}_{00} (\vv{k}) = \frac{2}{N} \sum_{\mu,\nu} e^{ i (\vv{k} - \vv{k}'_0) \cdot(\vv{r}_{\mu} - \vv{r}_{\nu})}, \label{eq:V022}\\
&V^{[2,2]}_{0\ell} (\vv{k}) = \mathcal{C}_{\mathcal{N}\mathcal{L}} \sum_{\mu} e^{i (\vv{k} - \vv{k}'_0)\cdot \vv{r}_{\mu}}  \sum_{\gamma = 1}^{\ell} S^{\ell}_{\gamma[2]}(\vv{k} - \vv{k}'_0). \label{eq:V0ell22}
\end{align}
We have defined the normalization coefficient $\mathcal{C}_{\mathcal{N}\mathcal{L}} = 1/\sqrt{\mathcal{N}\mathcal{L}}$ and the functions
\begin{align}
 &S^{\ell}_{\gamma[2]}(\vv{k} - \vv{k}'_0) = e^{- i(\vv{k} - \vv{k}'_0)\cdot \vv{r}_{\gamma(1)}}+  e^{- i(\vv{k} - \vv{k}'_0)\cdot \vv{r}_{\gamma(2)}}  \nonumber \\
 &- e^{- i(\vv{k} - \vv{k}'_0)\cdot \vv{r}_{\ell + 1(1)}} - e^{- i(\vv{k} - \vv{k}'_0) \cdot\vv{r}_{\ell + 1(2)}} \xrightarrow{\vv{k} \rightarrow \vv{k}'_0} 0.
 \label{eq:S_function}
 \end{align}
These expressions represent the leading order, while corrections are $O(1/N)$. This result shows that the symmetric excitations decouple from the non-symmetric states also in this case.
In particular, Eq.~\eq{V022} is sharply peaked at $\delta(\vv{k} - \vv{k}'_0),$ the detailed form depending on the atomic distribution. The functions \eq{S_function} vanish for $\vv{k} = \vv{k}'_0,$ causing suppression of \eq{V0ell22} close to that condition. The separation of symmetric and non-symmetric modes is related to a phase-matching condition as for the single excitation.

We write \eq{symmetric_double} for $\ell = 0$ taking into account only the leading coupling,
\begin{equation}
\frac{\partial \mathscr{E}_{0[2]}}{\partial t} = \sum_{\phi} - g^2_{\phi} \int_0^t dt' e^{-i \Delta\omega_{k} \tau} V^{[2,2]}_{00}(\vv{k}) \mathscr{E}_{0[2]} (t').
\label{eq:decay_double}
\end{equation}
From \eq{gamma_N}, we see that the amplitude for the double symmetric excitation decays with radiative width $2 \Gamma_N$ \cite{scullylaserphys2007}, with the photon emitted in the phase matched direction determined by wavevector $\vv{k}'_0$.

We consider the equation of motion for the amplitude of the single atomic excitation plus one photon:
\begin{align}
\frac{\partial \mathscr{E}^{\phi}_{\ell}}{\partial t}  = - g_{\phi} e^{-i \Delta\omega_k t} \sum_{\ell',\mu} e^{-i \vv{k}\cdot \vv{r}_{\mu}} \bra{\mathsf{E}_{\ell}} \hat{\sigma}^{ge}_{\mu} \ket{\mathsf{E}_{\ell'[2]}} \mathscr{E}_{\ell'[2]}.
\label{eq:single_from_double}
\end{align}
The coupling between the singly and doubly excited atomic states is fixed by the couplings
\begin{equation}
V^{[1,2]}_{\ell\ell'}(\vv{k}) = \sum_{\mu} e^{-i \vv{k}\cdot \vv{r}_{\mu}}  \bra{\mathsf{E}_{\ell}} \hat{\sigma}^{ge}_{\mu} \ket{E_{\ell'[2]}}.
\label{eq:singledouble_coupling}
\end{equation}
Among these matrix elements, we calculate those that contain at least one symmetric state,
\begin{align}
 V^{[1,2]}_{00}(\vv{k}) =&\; \frac{N-1}{\sqrt{N\mathcal{N}}} \sum_{\mu} e^{i (\vv{k}'_0 - \vv{k})\cdot \vv{r}_{\mu}}, \label{eq:V0012} \\
V^{[1,2]}_{0\ell}(\vv{k})  =&\;  C_{N\mathcal{L}} \sum_{\gamma = 1}^{\ell} S^{\ell}_{\gamma[2]}(\vv{k} - \vv{k}'_0), \label{eq:V0ell12}\\
 V^{[1,2]}_{\ell0}(\vv{k}) = &\; C_{\mathcal{N}\mathcal{L}} \sum_{\lambda = 1}^{\ell} \left[- S^{\ell}_{\lambda}(\vv{k} - \vv{k}'_0) \right].
 \label{eq:Vell012}
\end{align}
These expressions determine how a double atomic excitation decays to a single excitation with emission of a photon.
The dominant term is \eq{V0012}, which is peaked in the preferred direction $\vv{k} = \vv{k}'_0$.
The phase matched direction favors the coupling between the symmetric excitations with respectively two and one excited atoms.
This decay channel is separated from non-symmetric atomic states: the $S$ functions \eq{Sfunction1} and \eq{S_function}, for single and double excitations, appear in \eq{Vell012} and \eq{V0ell12}, suppressing the contribution of these states to the emission around $\vv{k}'_0$.
We specialize \eq{single_from_double} to the case $\ell' = 0$ corresponding to a double spin wave and reintroduce the coupling to the two-photon continua, giving
\begin{align}
\frac{\partial \mathscr{E}^{\phi}_\ell}{\partial t} = & - g_{\phi} e^{-i \Delta\omega_k t} V^{[1,2]}_{\ell0}(\vv{k}) \mathscr{E}_{0[2]} + \nonumber \\
&+ \sum_{\phi'}  g_{\phi'} e^{i \Delta\omega_{k'} t} V_{\ell G}(\vv{k}') \sqrt{\epsilon^{\phi\phi'}} G^{\phi\phi'}.
\label{eq:single_phiell}
\end{align}
The matrix elements $V_{\ell G}$ couples single atomic excitations and atomic ground state; we have already calculated the form for the symmetric state $V_{0G}$ in \eq{V0G}. The transition amplitudes to non-symmetric states are,
\begin{align}
V_{\ell G}(\vv{k}') &=  \frac{1}{\sqrt{\mathcal{L}}} \sum_{\alpha = 1}^{\ell} S_{\alpha}^{\ell}(\vv{k}' - \vv{k}'_0). \label{eq:VellG}
\end{align}
\label{eq:single_ground}
These results confirm the behavior found for a single excitation and show how the radiative interactions among atoms drive the decay of two spin waves (doubly excited symmetric timed Dicke state).

We can summarize the emission processes by the sequence:
\begin{equation}
\mathscr{E}_{0[2]} \xrightarrow{V^{[1,2]}_{00} \sim \delta(\vv{k} - \vv{k}'_0)} \mathscr{E}^{(\vv{k}'_0,\lambda)}_0 \xrightarrow{V_{0G} \sim \delta(\vv{k}' - \vv{k}'_0)} G^{(\vv{k}'_0,\lambda),(\vv{k}'_0,\lambda)}
\end{equation}
All other couplings are proportional to some function $f(\vv{k} - \vv{k}'_0) \rightarrow 0$ for $\vv{k} \rightarrow \vv{k}'_0$.

\subsection{Decay of a double spin wave}
Based on the approximations exploited above, we give the explicit temporal evolution of the amplitudes involved in the decay of a double spin wave. The initial conditions are:
\begin{equation}
 E_{\mu\nu}(0) = \sqrt{\frac{1}{\mathcal{N}}} e^{i \vv{k}'_0\cdot (\vv{r}_{\mu} + \vv{r}_{\nu})}, \quad \mu \neq \nu.
\end{equation}
By switching to the timed Dicke basis and through application of the WW approximation, the solution of \eq{decay_double} is
\begin{equation}
\mathscr{E}_{0[2]}(t) = e^{-\Gamma_N t} \mathscr{E}_{0[2]}(0),
\end{equation}
where $\Gamma_N$ is defined in \eq{gamma_N}.
In order to solve for $\mathscr{E}^{\phi}_{\ell}$ and $G^{\phi\phi'}$, we use \eq{single_phiell} and
\begin{align}
\sqrt{\epsilon^{\phi\phi'}} G^{\phi\phi'}= &- \int_0^t dt' \sum_{\ell} \left\{g_{\phi} e^{-i \Delta\omega_{k} t'}V_{G\ell}(\vv{k}) \mathscr{E}^{\phi'}_{\ell}(t') \right. \nonumber \\
&\left. + g_{\phi'} e^{-i \Delta\omega_{k'}  t'} V_{G\ell}(\vv{k}') \mathscr{E}^{\phi}_{\ell}(t')  \right\}.
\end{align}
We substitute this expression into \eq{single_phiell}, retaining the coupling between states with different single atomic excitations and the same photonic mode, $\mathcal{E}^{\phi}_{\ell}$ and $\mathcal{E}^{\phi}_{\ell'}$ , but discarding interaction terms between $\mathscr{E}^{\phi}_{\ell}$ and $\mathscr{E}^{\phi'}_{\ell},$ as discussed in appendix \ref{sec:appendixB}.
We obtain
\begin{align}
&\frac{\partial \mathscr{E}^{\phi}_{\ell}}{\partial t} = - g_{\phi} e^{- i \Delta\omega_k t} V^{[1,2]}_{\ell 0}(\vv{k}) \mathscr{E}_{0[2]}  \nonumber \\
&- \sum_{\phi',\ell'} g^2_{\phi'} V_{\ell G}(\vv{k}') V_{G \ell}(\vv{k}') \int_0^t dt' e^{- i \Delta\omega_{k'} \tau} \mathscr{E}^{\phi}_{\ell'}(t') .
\end{align}
Recall that the coupling $V^{[1,2]}_{\ell 0}$ is suppressed with respect $V^{[1,2]}_{00}$ and $V_{\ell G}$ is dominated by $V_{0G}$. We solve for $\mathscr{E}^{\phi}_{0}$ to the leading order. The WW approximation gives
\begin{equation}
\mathscr{E}^{\phi}_0(t) = g_{\phi} V^{[1,2]}_{00}(\vv{k})  \frac{e^{(-i\Delta\omega_k - \Gamma_N)t} - e^{-\frac{\Gamma_N}{2}t}}{i\Delta\omega_k + \Gamma_N/2} \mathscr{E}_{0[2]}(0).
\end{equation}
In the large ensemble limit the emission is strongly peaked in the phase-matched direction as previously stressed.

Finally, we solve for the two-photon amplitude $G^{\phi\phi'}$.
In the asymptotic limit $t \rightarrow \infty$, we obtain
\begin{equation}
G^{\phi\phi'} = \frac{1}{\sqrt{\epsilon^{\phi,\phi'}}}\left[\frac{g_{\phi}V^{[1,2]}_{00}(\vv{k})}{ i \Delta\omega_k + \Gamma_N/2} \frac{g_{\phi'}V_{0G}(\vv{k}')}{ i \Delta\omega_{k'} + \Gamma_N/2}\right] \mathscr{E}_{0[2]}(0). \label{eq:two_photon_ampl}
\end{equation}
By comparison with \eq{mode_ampl}, it is clear that the decay of the double spin wave results in a pair of photons emitted into the phase matched mode, \eq{phase_matched}:
\begin{equation}
\ket{\Psi} \xrightarrow{t \rightarrow \infty} (\hat{b}_{\vv{k}'_0}^{\dagger})^2\ket{G;0}.
\end{equation}

\subsection{Effect of atomic interactions}
We finally consider the effect of the phase shifts \eq{phase_shift} induced by two-body interactions for atoms in the metastable state $|r \rangle$.
The state $\ket{\mathsf{E}_{0[2]}}$ evolves into
\begin{equation}
\ket{\Phi} = \sqrt{\frac{1}{\mathcal{N}}} \sum_{\nu = 1}^{N}  \sum_{\mu > \nu} e^{i \vv{k}_0 \cdot(\vv{r}_{\mu} + \vv{r}_{\nu}) + i\Phi_{\mu\nu}} \ket{\mu,\nu}.
\end{equation}
We expand $\ket{\Phi}$ in terms of the symmetric and non-symmetric modes:
\begin{align}
\scalar{\mathsf{E}_{0[2]}}{\Phi}
&= \frac{1}{2 \mathcal{N}} \sum_{(\mu,\nu)} e^{i\Phi_{\mu\nu}}, \\
\scalar{\mathsf{E}_{\ell[2]}}{\Phi}
& = \sqrt{\frac{1}{\mathcal{L}\mathcal{N}}} \sum_{\jmath=1}^{\ell} \left[e^{i\Phi_{\jmath(1)\jmath(2)}} - e^{i\Phi_{\ell+1(1)\ell+1(2)}} \right].
\end{align}
In the limit $\Phi_{\mu\nu} \rightarrow 0$, we obtain the previous results: the amplitude for the symmetric timed Dicke state is $1$ while all the other terms vanish.
In the case that the phase shifts are large and broadly distributed, the amplitude of the symmetric state is quenched indicating the emission into the phase matched mode.
If we recall the state \eq{Psi0}, this quenching of the two-photon amplitude in the phase matched mode corresponds to the suppression of the spin wave correlation function, consistent with single photon emission in direction $\vv{k}'_0$.

\section{Multiply excited state}
\label{sec:n}
In this Section, we show that the results of Sec. \ref{sec:single} and \ref{sec:double} hold in the more general case of $n \ll N$ spin waves stored in the ensemble.
As for the double excitations, we assume that a laser pulse transfers the spin wave excitations from $\ket{r}$ to $\ket{e}$ before the emission takes place. In order to proceed, we refine the notation: we label the many-body atomic state through the indices of excited atoms $\ket{e_{\mu_1}....e_{\mu_n}} \rightarrow \ket{\mu_1...\mu_n}$. 

We consider the symmetric initial condition
\begin{equation}
E_{\mu_1...\mu_n}(0) = \sqrt{\binom{N}{n}^{-1}} e^{i \vv{k}'_0\cdot (\vv{r}_{\mu_1} + ...+ \vv{r}_{\mu_n})},
\end{equation}
while all the other amplitudes vanish at $t = 0$.

We introduce the timed Dicke basis for $m = 1,....,n$ material excitations:
\begin{subequations}
\begin{align}
\ket{\mathsf{E}_{0[m]}} =& \sqrt{\binom{N}{m}^{-1}} \frac{1}{m!} \sum_{(\mu_1...\mu_m)} e^{i \vv{k}'_0\cdot (\vv{r}_{\mu_1} + ...+ \vv{r}_{\mu_m})} \ket{\mu_1...\mu_m}, \\
\ket{\mathsf{E}_{\ell[m]}} =& \frac{1}{\sqrt{\mathcal{L}}} \sum_{\gamma = 1}^{\ell} \left[ e^{i \vv{k}'_0\cdot (\vv{r}_{\gamma(1)} + ...+ \vv{r}_{\gamma(m)})}\ket{\gamma(1)...\gamma(m)} \right. \nonumber \\
& \left. - e^{i \vv{k}'_0\cdot (\vv{r}_{\ell+1(1)} + ...+ \vv{r}_{\ell+1(m)})}\ket{\ell+1(1)...\ell+1(m)}  \right].
\end{align}
\end{subequations}
Here, following the case of double excitations, we define $\ell$ and $\gamma$ as indices for collections of $m$ different atoms, $\ell(1)$,$\ell(2)$,...,$\ell(m)$. The notation $[m]$ is a reminder that the state contains $m$ atomic excitations.

By using this basis and similar approximations to those described for the treatment of double excitations, we obtain an equation for the decay of the $n$th-excitation state,
\begin{align}
\frac{\partial \mathscr{E}_{\ell[n]}}{\partial t} =& - \sum_{\phi} g^2_{\phi} \sum_{\ell'} V_{\ell\ell'}^{[n,n]}(\vv{k}) \int_0^t dt' e^{-i \Delta\omega_k \tau} \mathscr{E}_{\ell'[n]}(t'),
\label{eq:decay_n}
\end{align}
which has the same structure as \eq{symmetric_double} for the double excitation.
Here, we define the radiative coupling among the timed Dicke states by,
\begin{equation}
V_{\ell\ell'}^{[n,n]}(\vv{k})  = \sum_{(\mu,\nu)} e^{i \vv{k}\cdot (\vv{r}_{\mu} - \vv{r}_{\nu})} \bra{\mathsf{E}_{\ell[n]}}\hat{\sigma}^{eg}_{\mu} \hat{\sigma}^{ge}_{\nu} \ket{\mathsf{E}_{\ell'[n]}},
\end{equation}
giving explictly,
\begin{align}
V^{[n,n]}_{00}(\vv{k}) = & \frac{n}{N} \sum_{(\mu,\nu)} e^{i(\vv{k} - \vv{k}'_0)\cdot(\vv{r}_{\mu} - \vv{r}_{\nu})} + O\left(\frac{1}{N}\right),
\label{eq:V00nn}
\end{align}
\begin{align}
V^{[n,n]}_{0\ell}(\vv{k})  = \frac{1}{\sqrt{\binom{N}{n}\mathcal{L}}} \sum_{\jmath = 1}^{\ell} &\left\{ \sum_{\mu} e^{i(\vv{k} - \vv{k}'_0) \cdot\vv{r}_{\mu}} S^{\ell}_{\jmath[n]}(\vv{k} - \vv{k}'_0) \right. \nonumber \\
& \left.+ O\left(\frac{1}{N}\right) \right\}, \label{eq:V0ellnn}
\end{align}
where we have generalized the definition of the functions \eq{Sfunction1} and  \eq{S_function} to
\begin{equation}
S^{\ell}_{\jmath[n]}(\vv{k} - \vv{k}'_0) = \sum_{s = 1}^{n} \left[ e^{i (\vv{k} - \vv{k}'_0) \cdot\vv{r}_{\jmath(s)}} - e^{i (\vv{k} - \vv{k}'_0) \cdot\vv{r}_{\ell+1(s)}} \right].
\label{eq:Sfunction}
\end{equation}
These expressions make evident also in this case the separation of the decay channels between the collective symmetric excitations and the rest of non-symmetric atomic states, as shown in \fig{mode_coupling} for the case $n = 1$. In fact, Eq.~\eq{V00nn} has a sharp peak for $\vv{k} = \vv{k}'_0$ where the couplings in \eq{V0ellnn} all vanish.
We obtain from \eq{decay_n} that the symmetric timed Dicke state $\ket{\mathsf{E}_{0[n]}}$ decays with rate $n\textrm{Re}(\Gamma_N)$.

By following the calculation for the double excitation case, we evaluate the decay process which leads from $n$ excited atoms to $n-1$ plus one photon, in order to characterize the emitted radiation.
By analogy with Eq.~\eq{single_from_double}, we define the matrix elements connecting the symmetric state $\ket{\mathsf{E}^{[n]}_{0}}$ to the basis with $n-1$ atomic excitations,
\begin{align}
&V^{[n-1,n]}_{\ell\ell'} = \sum_{\mu} e^{-i\vv{k} \cdot \vv{r}_{\mu}} \bra{\mathsf{E}_{\ell[n-1]}} \hat{\sigma}^{ge}_{\mu} \ket{\mathsf{E}_{\ell[n]}}.
\end{align}
Those couplings involving at least one symmetric state are,
\begin{align}
&V^{[n-1,n]}_{00}(\vv{k}) = \frac{\sqrt{n(N-n+1)}}{N} \sum_{\mu} e^{i (\vv{k}'_0 - \vv{k}) \cdot\vv{r}_{\mu}}, \label{eq:V00n-1}\\
&V^{[n-1,n]}_{0\ell}(\vv{k}) = \sqrt{\frac{1}{\binom{N}{n-1}\mathcal{L}}} \sum_{\jmath = 1}^{\ell} S^{\ell}_{\jmath[n]}(\vv{k}'_0 - \vv{k}),\label{eq:V0elln-1} \\
&V^{[n-1,n]}_{\ell0}(\vv{k}) = \sqrt{\frac{1}{\binom{N}{n}\mathcal{L}}} \sum_{\jmath = 1}^{\ell} [- S^{\ell}_{\jmath[n-1]}(\vv{k}'_0 - \vv{k})]. \label{eq:Vell0n-1}
\end{align}
These have similar properties to \eq{V0012}, \eq{V0ell12} and \eq{Vell012}. In fact, \eq{V00n-1} is peaked for $\vv{k} = \vv{k}'_0$ showing the predominant emission in the phase matched direction where the couplings \eq{V0elln-1}, \eq{Vell0n-1} are strongly suppressed because of the $S^{\ell}_{\jmath[n]}$ functions.
This indicates the first photon in the decay of the multiple spin wave will be emitted in direction $\vv{k}'_0$ and the atomic state will retain its collective symmetry.
This result for the $n \rightarrow (n-1)$ de-excitation, together with the results for the single and double excitations affirms by induction that the subspaces of symmetric and non-symmetric atomic states are separated in the radiative coupling with the quantized electromagnetic field. The argument is valid for any number of atomic excitations throughout the whole decay process. The final state of the system is then given by
\begin{equation}
\ket{\Psi} \xrightarrow{t \rightarrow \infty} (\hat{b}_{\vv{k}'_0}^{\dagger})^n\ket{G;0}.
\end{equation}
We stress that the calculation is valid only for a large number of atoms since we have neglected corrections of $O(n/N)$.

Two-body interactions that dephase the symmetric $n$ atom excitations in the metastable state will once again act to quench emission into the phase matched mode. The state resulting from the dephasing of $n$ spin waves, with pairwise phase shift $\Phi_{\mu\nu},$ is given by
\begin{flalign}
\ket{\Phi_{(n)}} = \sqrt{\frac{1}{\binom{N}{n}}}  \sum_{\mu_1>...>\mu_n} e^{i [\vv{k}'_0 \cdot \sum_{j} \vv{r}_{\mu_j} + \sum_{l,j} \Phi_{\mu_l\mu_j}]} \ket{\mu_1...\mu_n}.
\end{flalign}
The amplitude of the phase matched symmetric state is
\begin{equation}
\langle \mathsf{E}_{0[n]} \ket{\Phi_{(n)}} = \binom{N}{n}^{-1} \sum_{\mu_1>...>\mu_n} e^{i \sum_{l,j} \Phi_{\mu_l\mu_j}},
\end{equation}
which vanishes in the limit of large and broadly distributed phase shifts, except for the single excitation amplitude that results in single photon emission in the phase-matched direction.

\section{Conclusion}
\label{sec:conclusion}
We have described the radiative retrieval process of multiple spin waves stored in metastable Rydberg states of an atomic ensemble. We have shown that the decay from a weakly excited ensemble is strongly directional thanks to the enhanced coupling to a phase matched mode and suppression of the contribution of non-symmetric excitations.
The demonstration given is valid for an arbitrary number of excitations much smaller than the total number of atoms. The use of the timed Dicke basis allows to focus on the relevant phase matched mode and can be expressed in terms of the transition amplitudes between the timed Dicke states mediated by the electromagnetic field.
As expected, the radiative coupling to the phase matched mode contains information about the size of the ensemble and its fluctuations are related to the density and the shape of the cloud.
The analysis enables a mapping of atomic excitations into emitted photons. It indicates that symmetric atomic excitations, those created by laser driving, undergo phase matched emission. A pairwise interaction of metastable storage atoms, however, suppresses these symmetric amplitudes and quenches multiphoton emission into the phase matched mode, leaving only the unperturbed single photon emission process. This insight may have application to fast single photon sources based on cold atomic ensembles.

We acknowledge financial support from NSF and AFOSR.
We thank Y. Dudin, A. Kuzmich, H.H. Jen and S.D. Jenkins for useful discussions.

\appendix

\section{Review of the theory for a single excitation}
\label{sec:appendixA}

The case of a single excitation stored into an atomic gas has received extensive attention in the recent years. Here, we limit the discussion to review the relevant results \cite{scullylaserphys2007,scullyPRL2009} in a way appropriate for generalization to multiple excitations. We also use this example to show how to eliminate the dynamics of the metastable state $\ket{r},$ thus reducing significantly the complexity of the problem for the case of multiple excitations.
If a single excitation is present in the system, the general state may be written as \eq{single_state}.
The equations of motion for the amplitudes of the different basis states are
\begin{subeqnarray}
i \hbar \frac{\partial G^{\phi}}{\partial t} & = & -i \hbar \sum_{\mu} g_{\phi\mu} e^{-i \vv{k} \cdot \vv{r}_{\mu}} e^{-i \Delta\omega_k t} E_{\mu}, \slabel{eq:single_G} \\
i \hbar \frac{\partial E_{\mu}}{\partial t} & = & \frac{\hbar \Omega_L}{2} e^{-i \vv{k}_L \cdot \vv{r}_{\mu}} e^{-i \Delta\omega_L t} R_{\mu}  \nonumber \\
                                                                    && + i \hbar \sum_{\phi} g_{\phi\mu} e^{i \vv{k}\cdot \vv{r}_{\mu}} e^{i \Delta\omega_k t} G^{\phi} \slabel{eq:single_Emu}, \\
i \hbar \frac{\partial R_{\mu}}{\partial t} & = & \frac{\hbar \Omega_L}{2} e^{i \vv{k}_L \cdot \vv{r}_{\mu}} e^{i \Delta\omega_L t} E_{\mu}. \slabel{eq:single_Rmu}
\label{eq:single_exc}
\end{subeqnarray}
Here, we introduce the notation $\Delta\omega_k = \omega_{eg} - \omega_k$ and $\Delta\omega_L = \omega_{re} - \omega_L$.
We integrate \eq{single_Rmu} and \eq{single_G} and we substitute the results in \eq{single_Emu}:
\begin{align}
& \frac{\partial E_{\mu}}{\partial t} = - \frac{ \Omega_L(t)}{4} \int_{0}^{t} dt'  e^{i\Delta\omega_L \tau} \Omega_L(t') E_{\mu}(t') \nonumber \\
& - \sum_{\phi} g^2_{\phi} \int_{0}^{t} dt'  e^{-i\Delta\omega_k \tau} \sum_{\nu} e^{i \vv{k}\cdot (\vv{r}_{\mu} - \vv{r}_{\nu})} E_{\nu}(t').
                                            \label{eq:WW}
\end{align}
Here, we assume that the coupling coefficient $g_{\phi\mu}$ is independent from $\mu$, which is equivalent to consider all the atomic dipoles aligned in a given direction; this model can be realized by properly choosing the external fields used for excitation and trapping of the atoms in order to allow for a single transition to be active. We also define $\tau = t' - t$.
We substitute the sum over the e.m. modes with an integral, $\sum_{\phi = (\vv{k},\lambda)} \rightarrow \sum_{\lambda = 1,2} V/(2\pi^3) \int_{0}^{2\pi} d\alpha \int_{0}^{\pi} d\theta \sin\theta \int_{0}^{\infty} d\omega \omega^2/c^3$, where the angles $(\alpha,\theta)$ characterize the direction of the wavevector of the photon and  $\lambda$ labels its polarization; the sum over the polarization for the coupling constant gives $\sum_{\lambda=1,2} \hbar g^2_{\phi} \rightarrow  \omega (1 - \cos^2\theta) \; d^2_{eg}/(2 \epsilon_0 V)$.
By using the Wigner-Weisskopf approach \cite{scullyquantopt}, we obtain \eq{single_decay} from \eq{WW}.
In this latter equation, the function $f$ is defined as \cite{LehmbergPRA1}
\begin{equation}
f^{k}_{\mu\nu} = \frac{3}{8\pi} \int_{0}^{2\pi} d\alpha \int_{0}^{\pi} d\theta \sin\theta (1 - \cos^2\theta) e^{i \vv{k}\cdot (\vv{r}_{\mu} - \vv{r}_{\nu})},
\label{eq:f_function}
\end{equation}
and it depends on the relative distance between particles $\mu$ and $\nu$ as well as on the wavevector of the emitted photon.
Equation \eq{single_decay} corresponds to the expression derived in literature \cite{LehmbergPRA1,scullypra2010,manassahPRA2010}: it is possible to analytically solve it in terms of Bessel functions for the case of a scalar photon. Since we are interested in the decay of a spin wave, we proceed in a different way by introducing the timed Dicke basis for the atomic wavefunction.

\subsection{Timed Dicke basis for single excitation}
The timed Dicke basis for a single excitation in the ensemble is defined as follows \cite{scullylaserphys2007}
\begin{subequations}
\begin{flalign}
&\ket{\mathsf{R}_0}= \frac{1}{\sqrt{N}} \sum_{\mu = 1}^{N} e^{i\vv{k}_0\cdot\vv{r}_{\mu}} \ket{s_{\mu}},\\
&\ket{\mathsf{R}_{\ell}} = \frac{1}{\sqrt{\mathcal{L}}} \sum_{\mu = 1}^{\ell} \left[  e^{i\vv{k}_0\cdot\vv{r}_{\mu}} \ket{s_{\mu}} - e^{i \vv{k}_0 \cdot\vv{r}_{\ell+1}} \ket{s_{\ell+1}} \right];
\end{flalign}
\label{eq:scully_single}
\end{subequations}
here  $\ell \in [1,N-1]$, $\mathcal{L} = \ell(\ell+1)$.  If the excitation is in the level $\ket{e}$, we substitute $\mathsf{R} \rightarrow \mathsf{E}$ and we use the wavevector $\vv{k}'_0 = \vv{k}_0 - \vv{k}_L$.
In general we refer to the $\ell = 0$ state as the symmetric state since the excitation is shared between all the atoms with equal probability, while all the other basis vectors are non-symmetric in this sense.
We rewrite the state \eq{single_state} in terms of this new basis
\begin{equation}
\ket{\Psi} = \sum_{\phi} G^{\phi} \ket{G;\phi} + \sum_{\ell = 0}^{N-1} \left( \mathscr{E}_{\ell} \ket{\mathsf{E}_{\ell};0} + \mathscr{R}_{\ell} \ket{\mathsf{R}_{\ell};0}\right) .
\end{equation}

We write the equations for the amplitudes of the new basis states:
\begin{subequations}
\begin{align}
\frac{\partial G^{\phi}}{\partial t}  =&\; - g_{\phi} e^{-i\Delta\omega_k t} \sum_{\mu,\ell}  e^{-i \vv{k} \cdot\vv{r}_{\mu}}  \bra{G} \hat{\sigma}^{ge}_{\mu} \ket{\mathsf{E}_{\ell}} \mathscr{E}_{\ell}, \label{eq:scully_G} \\
 \frac{\partial \mathscr{E}_{\ell}}{\partial t}  = &\;-i\frac{ \Omega_L}{2} e^{-i\Delta\omega_L t}  \mathscr{R}_{\ell} \nonumber \\
&+    \sum_{\mu,\phi} g_{\phi} e^{i\Delta\omega_k t}  e^{i \vv{k} \cdot\vv{r}_{\mu}} \bra{\mathsf{E}_{\ell}} \hat{\sigma}^{eg}_{\mu} \ket{G} G^{\phi}, \label{eq:scully_E} \\
\frac{\partial \mathscr{R}_{\ell}}{\partial t} = &\;- i \frac{ \Omega_L}{2} e^{i\Delta\omega_L t} \mathscr{E}_{\ell}.  \label{eq:scully_R}
\end{align}
\end{subequations}
We integrate \eq{scully_G} and substitute the formal result for $G^{\phi}$ into \eq{scully_E} to determine the coupling between the different timed Dicke states mediated by the field.
We analyze the behavior of the transition amplitudes for ensemble emission and re-absorption between the symmetric timed Dicke state $\ell = 0$ with itself, $\ell'=0$, and with the rest of the basis, $\ell' \neq 0$:
\begin{align}
&\bra{\mathsf{E}_0}\hat{\sigma}^{eg}_{\mu} \ket{G}\bra{G}\hat{\sigma}_{\nu}^{ge}\ket{\mathsf{E}_0} = \frac{1}{N} e^{i \vv{k}_0' \cdot (\vv{r}_{\nu} - \vv{r}_{\mu})}, \\
&\bra{\mathsf{E}_0}\hat{\sigma}^{eg}_{\mu} \ket{G}\bra{G} \hat{\sigma}_{\nu}^{ge} \ket{\mathsf{E}_{\ell \neq 0}} = \frac{1}{\sqrt{N}} \sum_{\alpha = 1}^{N}e^{- i \vv{k}_0' \cdot \vv{r}_{\alpha}} \delta_{\alpha \mu}  \nonumber \\
 &\times \frac{1}{\sqrt{\ell(\ell + 1)}} \sum_{\beta = 1}^{\ell}  \left[e^{i \vv{k}_0' \cdot \vv{r}_{\beta}} \delta_{\nu\beta} - e^{i \vv{k}_0' \cdot\vv{r}_{\ell + 1}} \delta_{\nu\ell+1} \right].
\end{align}
By using these matrix elements, we are lead to \eq{decayE0}.

\subsection{Retrieval of a single spin wave}
As initial condition, we assume a timed Dicke state stored in the metastable state with wavevector $\vv{k}_0$
\begin{equation}
 R_{\mu}(0) = \frac{1}{\sqrt{N}} e^{i \vv{k}_0 \cdot\vv{r}_{\mu}},
\label{eq:sw_ampl}
\end{equation}
while all the other amplitudes vanish at the intial time.
We may model the retrieval of this spin wave by means of a $\pi$-pulse from the laser field $\Omega_L$ followed by the  decay from the state $\ket{e}$.
We consider a resonant pulse, $\omega_L = \omega_{re}$.
Furthermore, we assume a pulse whose duration $T \approx \pi/\overline{\Omega}_{L}$, where $\overline{\Omega}_L$ is the temporal average of the laser Rabi frequency, is much smaller than the ensemble decay rate, $T \ll 1/\textrm{Re}(\Gamma_N)$.  Within these conditions and neglecting the weak coupling to non-symmetric states, the solution of \eq{decayE0} may be written as
\begin{equation}
\mathscr{E}_{0}(t) = A \; \sin\left[\int_0^t \frac{\Omega_L(t')}{2} dt' \right] e^{-\frac{\Gamma_N}{2}t};
\label{eq:E0t}
\end{equation}
the factor $A$ is set by the intial condition $A = \mathscr{R}_0(0) = 1$.
We define the phase $\beta(t) = \int_{0}^{t} dt' \Omega_L(t')/2$, and we calculate the derivative in time of the function \eq{E0t}:
\begin{align}
&\frac{\partial \mathscr{E}_{0}}{\partial t} = - \frac{\Omega_L(t)}{2} \int_0^t dt' \frac{\Omega_L(t')}{2} \mathscr{E}_0(t')  \nonumber\\
&- \frac{\Omega_L(t)}{2} \frac{\Gamma_N}{2} \int_0^t dt' A \cos\left[\beta(t') \right] e^{-\frac{\Gamma_N}{2}t'} -\frac{\Gamma_N}{2} \mathscr{E}_0(t),
\end{align}
where we have used the fact that $\partial_t \beta = \Omega_L(t)/2$.
We remark that the approximation holds for
\begin{equation}
\frac{\Omega_L(t)}{2} \frac{\Gamma_N}{2} \int_0^t dt' A \cos\left[\beta(t') \right] e^{-\frac{\Gamma_N}{2}t'} \approx \frac{\Gamma_N}{2}T    \ll 1,
\label{eq:temporal_approx}
\end{equation}
which is exactly the regime we are assuming. It is also possible to allow for a temporal dependence of the coefficient $A$ which provides an integro-differential equation for this quantity which turns out to be constant in the case \eq{temporal_approx}.
The temporal evolution is then separated in two parts: for $t \lesssim T$, the amplitude for the symmetric excitation in the intermediate state $\mathcal{E}_0$ grows following the laser pulse, while for $t \gg T$ the coupling to the continuum of the radiation modes transfers the excitation to the field. In this temporal range, we can approximate $\sin(\beta) \approx 1$ and the wavefunction asymptotically coincides with the e.m. field amplitude \eq{mode_ampl}.

\section{Wigner-Weisskopf approach for multiple excitations}
\label{sec:appendixB}

We discuss the two approximations used in the application of WW theory to multiple excitations.
We consider the case of double excitations, since it contains all the ingredients necessary to this demonstration.
In order to show the validity of the approximations, we use the set of equations \eq{double_eq_mot}.
We first focus on the coupling between the continua of one- and two-photon states.
We integrate Eq. \eq{double_exc} and we substitute the result in \eq{atom_photon} to obtain:
\begin{widetext}
\begin{align}
\frac{\partial E^{\phi}_{\mu}}{\partial t}  = & - \sum_{\nu\neq\mu} g_{\phi} e^{-i\vv{k}\cdot\vv{r}_{\nu}} e^{-i \Delta\omega_k \tau} E_{\mu\nu}
 - \sum_{\phi'} \left[ g^{2}_{\phi'} \int_{0}^{t} dt' e^{i \Delta\omega_{k'} (t-t')} \sum_{\nu=1}^{N} e^{- i \vv{k}' \cdot (\vv{r}_{\mu} - \vv{r}_{\nu})} E^{\phi}_{\nu} \right. \nonumber \\
 & \left. +   g_{\phi'}  g_{\phi}   e^{i \Delta\omega_{k'} t}  \int_{0}^{t} dt' e^{i \Delta\omega_{k} t'} \sum_{\nu=1}^{N} e^{- i( \vv{k}' \cdot \vv{r}_{\mu} -  \vv{k} \cdot \vv{r}_{\nu})} E^{\phi'}_{\nu}  \right].
\label{eq:approx_1}
\end{align}
\end{widetext}

In the spirit of WW, we substitute the sum over the e.m. modes $\phi'$ with an integral.
This integral can be separated in a frequency and an angular part: the former one is used to perform the Markov approximation. We see the difference between the two terms in the bracket. In fact, the coupling to $E^{\phi}_{\nu}$ depends on the phase factor $e^{i \Delta\omega_{k'} (t - t')}$ and, together with the integral over $\omega' = ck'$, it gives a function $\delta(t - t')$. The coupling to the amplitudes $E^{\phi'}_{\nu}$ contains instead $e^{i \Delta\omega_{k'} t}$ that results in $\delta(t)$: in the solution of the equation, the contribution of this term vanishes. We are thus allowed to discard the couplings $E^{\phi}_{\mu} \leftrightarrow E^{\phi'}_{\nu}$.
After application of WW, we are left with the equation
\begin{widetext}
\begin{align}
\frac{\partial E^{\phi}_{\mu}}{\partial t}  = & - \sum_{\nu\neq\mu} g_{\phi} e^{-i\vv{k}\cdot\vv{r}_{\nu}} e^{-i \Delta\omega_k (t-t')} E_{\mu\nu}(t) - \frac{\Gamma}{2} \sum_{\nu = 1}^{N} f^{k_{eg}}_{\mu\nu} E^{\phi}_{\nu}(t),
\label{eq:approx_1WW}
\end{align}
\end{widetext}
where $\Gamma$ is the single atom decay rate and we use the definition of the $f$ function \eq{f_function}.

A similar argument is valid to check the second approximation which consists of neglecting the terms depending on $G^{\phi\phi'}$ when we substitute \eq{atom_photon} in \eq{double_atoms}.
We consider the expression in \eq{approx_1WW}. We formally integrate the expression and we substitute the result in \eq{double_atoms}, which yields
\begin{widetext}
\begin{align}
\frac{\partial E_{\mu\nu}}{\partial t}  = & - \sum_{\phi} g^2_{\phi} \int_{0}^t dt' e^{-i \Delta\omega_k (t-t')} e^{-\frac{\Gamma}{2}(t - t')}\left[\sum_{\sigma \neq \nu} e^{i \vv{k}\cdot (\vv{r}_{\mu} - \vv{r}_{\sigma})} E_{\sigma\nu}(t') + \sum_{\sigma \neq \mu} e^{i \vv{k}\cdot (\vv{r}_{\mu} - \vv{r}_{\sigma})} E_{\sigma\mu}(t') \right] \nonumber \\
&- \sum_{\phi} g_{\phi} e^{i \Delta\omega_k t} \int_0^t dt'  e^{-\frac{\Gamma}{2}(t - t')} \frac{\Gamma}{2} \left[ e^{i \vv{k}\cdot \vv{r}_{\mu}} \sum_{\sigma \neq \nu} f^{k_{eg}}_{\nu\sigma} E_{\sigma}^{\phi}(t') + e^{i \vv{k}\cdot \vv{r}_{\nu}} \sum_{\sigma \neq \mu} f^{k_{eg}}_{\mu\sigma} E_{\sigma}^{\phi}(t')\right].
\label{eq:23complete}
\end{align}
\end{widetext}
We notice that the term in the second line of this equation contains the same type of sum over the continuum of radiation modes as the last term in \eq{approx_1}. In fact, the integral over the modulus $k$ together with the oscillating term $e^{i \Delta\omega_{k} (t - t')}$ gives a $\delta(t-t')$ in the terms in the first line, while its combination with $e^{i \Delta\omega_{k} t}$ in the second line produces the function $\delta(t)$: these latter terms then vanish when we perform the Markov approximation within WW and they do not affect the decay of the double excitations.
It is possible to understand this effect in the following picture: the linewidth acquired by the states $E^{\phi}_{\mu}$ as a consequence of the coupling to the amplitudes $G^{\phi\phi'}$  is much smaller than the bandwidth of their continuum of states, responsible for the decay of $E_{\mu\nu}$. While studying the behavior of $E_{\mu\nu}$, it is then possible to discard the terms depending on $G^{\phi\phi'}$: this leads to Eq.\eq{23} in the main text.
Furthermore, it is straightforward to verify that these approximations allow to find the correct results for the case of $N = 2$ atoms and $n = 2$ excitations, which agree with the results shown in \cite{ernst1968}. These arguments extend in the same way to the multiply excited states.

\bibliography{references_Url}

\end{document}